# A Possibility of Gravity Control in Luminescent Materials

Fran De Aquino

physics/0109060

Maranhao State University,
Physics Department,
65058-970 S.Luis/MA, Brazil.
E-mail: deaquino@elo.com.br

## Abstract

It was demonstrated (gr-qc/9910036) that the gravitational and inertial masses are correlated by an adimensional factor, which depends on the incident ( or emitted)radiation upon the particle. There is a direct correlation between the radiation absorbed( or radiated) by the particle and its gravitational mass, independently of the inertial mass. Only in the absence of electromagnetic radiation the mentioned factor becomes equal to *one*. On the other hand, in specific electromagnetic conditions, it can be reduced, nullified or made negative. This means that there is the possibility of the gravitational masses can be reduced, nullified or made negative by means of electromagnetic radiation. This unexpected theoretical result was confirmed by an experiment using Extra-Low Frequency(ELF) radiation on ferromagnetic material (gr-qc/0005107). Recently another experiment using UV light on phosphorescent plastic have confirmed the phenomenon. We present a complete explanation for the alterations of the gravitational field in luminescent materials. This work indicates that the alterations of the gravitational field can be sufficiently strong to invert the gravity on *luminescent materials*.

## Introduction

It is known that the physical property of mass has two distinct aspects, *gravitational mass* $m_g$ and *inertial mass* $m_i$. Gravitational mass produces and responds to gravitational fields. It supplies the mass factors in Newton's famous inverse-square law of gravity($F_{12}=Gm_{g1}m_{g2}/r_{12}^2$). Inertial mass is the mass factor in Newton's 2nd Law of Motion (F=$m_i$a**)**.

In a previous paper[1] we have shown that the *gravitational mass* and the *inertial mass* are correlated by an adimensional factor, which depends on the incident radiation upon the particle. It was shown that only in the absence of electromagnetic radiation this factor becomes equal to 1 and that, in specific electromagnetic conditions, it can be reduced, nullified or made negative.

The general expression of correlation between gravitational mass $m_g$ and inertial mass $m_i$, is given by

$$m_g = m_i - 2\left\{\sqrt{1+\left\{\frac{q}{m_i c}\right\}^2} - 1\right\}m_i \quad (1)$$

where the *momentum* $q$, according to the Quantum Mechanics, is given by

$$q = N\hbar k = N\hbar\omega/(\omega/k) = U/(dz/dt) =$$
$$= U/v \quad (2)$$

where $U$ is the electromagnetic energy absorbed (or emitted) by the particle; $v$ is the velocity of the incident (or emitted) radiation. It can be shown that

$$v = \frac{c}{\sqrt{\frac{\varepsilon_r \mu_r}{2}\left(\sqrt{1+(\sigma/\omega\varepsilon)^2}+1\right)}} \qquad (3)$$

where ; $\varepsilon$, $\mu$ and $\sigma$, are the electromagnetic characteristics of the outside medium around the particle in which the incident radiation is propagating ( $\varepsilon = \varepsilon_r \varepsilon_0$ where $\varepsilon_r$ is the *relative electric permittivity* and $\varepsilon_0 = 8.854 \times 10^{-12} F/m$; $\mu = \mu_r \mu_0$ where $\mu_r$ is the *relative magnetic permeability* and $\mu_0 = 4\pi \times 10^{-7} H/m$ ). For an *atom* inside a body, the incident (or emitted) radiation on this atom will be propagating inside the body, and consequently, $\sigma = \sigma_{body}$, $\varepsilon = \varepsilon_{body}$, $\mu = \mu_{body}$.

By substitution of Eqs.(2) and (3) into Eq.(1), we obtain

$$m_g = m_i - 2\left\{\sqrt{1+\left[\frac{U}{m_i c^2}\sqrt{\frac{\varepsilon_r \mu_r}{2}\left(\sqrt{1+(\sigma/\omega\varepsilon)^2}+1\right)}\right]^2} - 1\right\}m_i$$

$$= m_i - 2\left\{\sqrt{1+\left[\frac{U}{m_i c^2}n_r\right]^2} - 1\right\}m_i \qquad (4)$$

In the equation above $n_r$ is the *refractive index*, which is given by:

$$n_r = \frac{c}{v} = \sqrt{\frac{\varepsilon_r \mu_r}{2}\left(\sqrt{1+(\sigma/\omega\varepsilon)^2}+1\right)} \qquad (5)$$

$c$ is the speed in vacuum and $v$ is the speed in medium.

If the incident (or emitted) radiation is *monochromatic* and has frequency $f$, we can put $U = nhf$ in Equation (4), where $n$ is the number of incident (or radiated) photons on the particle of mass $m_i$. Thus we obtain

$$m_g = m_i - 2\left\{\sqrt{1+\left[\frac{nhf}{m_i c^2}n_r\right]^2} - 1\right\}m_i \qquad (6)$$

In that case, according to the *Statistical Mechanics*, the calculation of $n$ can be made based on the well-known method of *Distribution Probability*. If all the particles inside the body have the same mass $m_i$, the result is

$$n = \frac{N}{A}a \qquad (7)$$

where $N/A$ is the average density of incident (or emitted) photons on the body; $a$ is the area of the surface of a particle of mass $m_i$ from the body.

Obviously the power $P$ of the incident radiation must be $P = Nhf/\Delta t = Nhf^2$, thus we can write $N = P/hf^2$. Substitution of $N$ into Eq.(7) gives

$$n = \frac{a}{hf^2}\left(\frac{P}{A}\right) = \frac{a}{hf^2}D \qquad (8)$$

where $D$ is the *power density* of the incident (or emitted) radiation. Thus Eq.(6) can be rewritten in the following form:

$$m_g = m_i - 2\left\{\sqrt{1+\left[\frac{aD}{m_i cvf}\right]^2} - 1\right\}m_i \qquad (9)$$

For $\sigma >> \omega\varepsilon$ Eq.(3) reduces to

$$v = \sqrt{\frac{4\pi f}{\mu\sigma}} \qquad (10)$$

By substitution of Eq.(10) into Eq.(9) we obtain

$$m_g = m_i - 2\left\{\sqrt{1+\left[\frac{aD}{m_i c}\sqrt{\frac{\mu\sigma}{4\pi f^3}}\right]^2} - 1\right\}m_i \quad (11)$$

This equation shows clearly that, *atoms* (or *molecules*) can have their *gravitational masses* strongly reduced by means of Extra-Low Frequency (ELF) radiation.

We have built a system (called System G) to verify the effects of the ELF radiation on the gravitational mass of a body. In the system G, a 60Hz frequency radiation was produced by an ELF antenna. A thin layer of *annealed pure iron* around the antenna ( toroid form ) have absorbed all the ELF radiation.

In this annealed iron toroid $\mu_r = 25000$ ($\mu = 25000\mu_0$) and $\sigma = 1.03 \times 10^7 S/m$. The power density $D$ of the incident ELF radiation reach approximately 10Kw/m$^2$. By substitution of these values into Eq.11) it is easy to conclude the obtained results.

The experimental setup and the obtained results were presented in a paper[2].

The experiment above mentioned have confirmed that the general expression of correlation between gravitational mass and inertial mass (Eq.4) is correct. In practice, this means that the gravitational forces can be reduced, nullified and *inverted* by means of electromagnetic radiation.

Recently another experiment using UV light on *phosphorescent* materials have confirmed the phenomenon[3].

In this paper we present a complete explanation for the alterations of the gravitational field in *luminescent* ( photo, electro, thermo and tribo )materials. It was shown that the alterations of the gravitational field can be sufficiently strong to invert the gravity on *luminescent* materials.

## 1. Theory

When the material is *luminescent*, the radiated photons number $n$, radiated from the electrons, cannot be calculated by the Eq.(7), because, according to the quantum *statistical mechanics*, they are *undistinguishable* photons of varying frequencies and consequently follow the *Einstein-Bose statistics*. In that case, as we know, the number $n$ ( $n$ is the number of photons with frequency between $f$ and $f+\Delta f$ ) will be given by

$$n = \left(\frac{8\pi V f^2}{v^3}\right)\frac{1}{e^{E/m_i c^2}-1}\int_f^{f+\Delta f} df \cong$$

$$\cong \left(\frac{8\pi V}{v^3}\right)\frac{1}{e^{E/m_i c^2}-1} f^2 \Delta f$$

Thus, assuming $\Delta f \cong 1Hz$ (*quasi*-mono chromatic ) we obtain

$$n \cong \left(\frac{8\pi V f^2}{v^3}\right)\frac{1}{e^{E/m_i c^2}-1} \cong \left(\frac{8\pi V f^2}{v^3}\right)\frac{1}{e^{q/m_i c}-1} \cong$$

$$\cong \left(\frac{8\pi V}{v\lambda^2}\right)\frac{1}{e^{\lambda_c/\lambda}-1} \quad (12)$$

where $\lambda_c = h/m_i c$ is the *Compton wavelength* for the particle of mass $m_i$ and $\lambda$ is the *average* wavelength of the light emitted from the particle ; $V$ is the volume of the body which contains the particle.

By substitution of Eq.(12) into Eq.(6) we obtain

$$m_g = m_i - 2\left\{\sqrt{1+\left[\left(\frac{8\pi V}{v\lambda^2}\right)\frac{\lambda_c/\lambda}{e^{\lambda_c/\lambda}-1}\right]^2} - 1\right\}m_i \quad (13)$$

For $\sigma \ll \omega\varepsilon$ the Eq.(3) reduces to

$$v = \frac{c}{\sqrt{\varepsilon_r \mu_r}} \quad \text{and} \quad \lambda = \frac{c}{f\sqrt{\varepsilon_r \mu_r}}$$

Consequently Eq.(13) can be rewritten In the following form

$$m_g = m_i - 2\left\{\sqrt{1+\left[\left(\frac{8\pi V}{c^3}\right)f^2 n_r^3 \frac{\lambda_c/\lambda}{e^{\lambda_c/\lambda}-1}\right]^2} - 1\right\}m_i \quad (14)$$

For *electrons* $m_i = m_e = 9.11\times10^{-31} kg$ and $\lambda_{c(electrons)} = 2.42\times10^{-12} m$. For atoms $\lambda_{c(atoms)} \ll \lambda_{c(electrons)}$. On the other hand, if

$$\lambda \gg \lambda_c \Rightarrow \frac{\lambda_c/\lambda}{e^{\lambda_c/\lambda}-1} \approx 1$$

Then Eq.(14) reduces to

$$m_g = m_i - 2\left\{\sqrt{1+\left[\left(\frac{8\pi V}{c^3}\right)f^2 n_r^3\right]^2} - 1\right\}m_i \quad (15)$$

In the *Hardeman* experiment [3]

$$\left(\frac{8\pi V}{c^3}\right)f^2 n_r^3 \approx 0.658$$

Consequently, from Eq.(15) we obtain for the electrons of the *luminescent* material:

$$m_{g(electrons)} \approx 0.605 m_{i(electons)}$$

This means a 39.5% reduction in gravitational masses of the electrons of the atoms of the *phosphorescent* material. Thus, the *total* reduction in gravitational mass of the *phosphorescent* material will be given by

$$\frac{m_e - 0.605\,m_e}{m_e + m_p + m_n}\times 100\% \approx -0.011\%$$

Exactly the value obtained in the *Hardeman* experiment.

Now we will calculate the power of UV radiation, necessary to produce the reduction of weight, detected by Hardeman in the phosphorescent material.

According to the Quantum *Statistical Mechanics* the *gas of photons* inside a *luminescent material* has a *average* number of photons $N$, where

$$N = \frac{1}{e^{E/m_i c^2}} = \frac{1}{e^{\lambda_c/\lambda}-1} \quad (16)$$

This means that the UV *power P* should have the following value

$$P = Nhf^2 = \frac{hc^2}{\lambda^2(e^{\lambda_c/\lambda}-1)} \quad (17)$$

For $\lambda = 365\,nm$ (*UV light*). The equation above gives

$$P \approx 68\,W$$

From the Electrodynamics we know that a radiation with frequency $f$ propagating within a material with electromagnetic characteristics $\varepsilon$, $\mu$ and $\sigma$ has the amplitudes of its waves decreased of $e^{-1}=0.37$ (37%) when it penetrates a distance $z$, given by

$$z = \frac{1}{\omega\sqrt{\tfrac{1}{2}\varepsilon\mu\left(\sqrt{1+(\sigma/\omega\varepsilon)^2}-1\right)}} \quad (18)$$

The radiation is totally absorbed if it penetrates a distance $\delta \cong 5z$.

Thus, if we put under UV radiation ($\lambda$=365nm , > 68w) a sheet of phosphorescent plastic with $\sigma\ll 1S/m$ and $\varepsilon>\varepsilon_0$ ; $\mu>\mu_0$, the Eq. above tell us that $z \gg 5mm$. Consequently, we can assume that the UV radiation at 365nm has a good penetration within the plastic sheet above ( thickness = 2mm).

On the other hand, if we assume that the sheet has an index of refraction $n_r$~1, thus , according to Eq.(15), for $f$~$6\times10^{14}$Hz (green light radiated from the sheet), the gravitational mass of the *electrons* of the sheet will be NEGATIVE and given by

$$m_g = m_e - 2\left\{\sqrt{1+\left\{\left(\frac{8\pi V}{c^3}\right)f^2 n_r^3\right\}^2} - 1\right\}m_e \cong$$

$$\cong -335.1 m_e \qquad (19)$$

Thus, the *total* reduction in **gravitational mass of the sheet of** *phosphorescent* **material** will be given by

$$\frac{m_e - 335.1 m_e}{m_e + m_p + m_n} \times 100\% \approx -9.1\%$$

If $V = 2m \times 1.36m \times 0.002m = 0.00544 m^3$ we will have a reduction of approximately **100%**.

## 2. The Gravitational Motor

For $\sigma \gg \omega\varepsilon$, as we have seen, Eq.(3) reduces to Eq.(10), i.e.,

$$v = \sqrt{\frac{4\pi f}{\mu\sigma}}$$

Consequently Eq.(13), for $\lambda \gg \lambda_c$, can be rewritten in the following form,

$$m_g = m_i - 2\left\{\sqrt{1+\left\{\left(\frac{8\pi V}{c^3}\right)f^2 n_r^4\right\}^2} - 1\right\}m_i \qquad (20)$$

The difference between Eq.(20) and Eq.(15) is in exponent of the *index of refraction* $n_r$.

Both Eq.(15) and Eq.(20) tell us that *luminescent materials* with *high refractive indices* can be very efficients in gravity control technology.

In the particular case of the Gravitational Motor (presented in a previous paper[4]) these materials can simplify its construction.

Let us consider figure 1 where we present a new design for the Gravitational Motor based on *electroluminescent* materials.

The *average* mechanical *power* $P$ of the motor is

$$P = T\omega = (F\bar{r})\omega = (m_g g)\bar{r}\left(\frac{g}{\bar{r}}\right)^{\frac{1}{2}} =$$

$$= m_g \sqrt{g^3 \bar{r}} \qquad (21)$$

where $\bar{r} = R - (R_0 + \Delta r)$, (see Fig.1-a) and $m_g$ is the *gravitational mass* of the *electroluminescent* material inside the left-half of the rotor ( when NEGATIVE, obviously ) ( see rotor in Fig.1). It is easy to show that $m_g$ may be written in the form

$$m_g \cong -\left[\frac{N(Km_e)}{N(m_e + m_p + m_n)}\right]m_i \qquad (22)$$

for $Km_e > m_e + m_p + m_n; K > 3666.3$

where $m_p = m_n = 1.67 \times 10^{-27} kg$ are the masses of the proton and neutron respectively and $K$, in agreement with Eq.(20), is given by

$$K \approx 2\left[\frac{8\pi V f^2}{c^3}n_r^4\right] \qquad (23)$$

By substitution of Eq.(23) into Eq.(22) we obtain

$$m_g \cong -\left\{2\left[\frac{8\pi V f^2}{c^3}n_r^4\right]\frac{m_e}{m_e + m_p + m_n}\right\}m_i \qquad (24)$$

But the *electroluminescent* (EL) *material* of the rotor is divided in disks to reduce the gravitational pressure on them (see Fig.1-b). These disks (organic luminescent material) are between electrodes and submitted to suitable alternating voltage $\Delta V$ to emit *blue* light (frequency $f = 6.5 \times 10^{14} hz$). Thus, according to Eq.(24), the gravitational mass $m_g^1$ of *one* EL *disk*, (with volume $V = \pi R_0^2 \xi$ where

$R_0, \xi$ are respectively, the radius and the thickness of the EL disk), is given by

$$m_g^1 \cong -\left\{\left[\frac{16\pi\left(\pi R_0^2 \xi\right)^2 f^2}{c^3} n_r^4\right] \frac{m_e}{m_e + m_p + m_n}\right\}\rho \quad (25)$$

$$K \approx 2\left[\frac{8\pi V f^2}{c^3} n_r^4\right] = \frac{16\pi\left(\pi R_0^2 \xi\right) f^2}{c^3} n_r^4$$

For example, if the rotor has $R = 627mm; L = 1350mm$ and the EL disks:
$R_0 = 190mm; \rho \cong 800 kg/m^3; \xi = 45mm;$

$\chi = 0.2mm$ and $n_r \cong 1$

then the *gravitational mass* of each EL disk (ON) is

$$m_g^1 \cong -4.4 kg \quad ; \quad K \cong 4014$$

If the left-half of the rotor has $N_l = 2\frac{L-\chi}{\xi+\chi} = 60$ EL disks(ON), then the *total* gravitational mass $m_g$ is

$$m_g = N_l m_g^1 = 264 kg$$

Thus, according to Eq.(21) the power of the motor is

$$P = (264)\sqrt{(9.8)^3(0.627-(0.190+0.002))} \cong$$

$$\cong 5.3 Kw \cong 7 HP$$

A electric generator coupled at this motor can produce for one month an amount of electric energy $W$ given by

$$W = P.\Delta t = (5300w)(2.59 \times 10^6 s) =$$

$$= 1.4 \times 10^{10} j \cong 3800 Kwh$$

It is important to note that if $n_r \cong 2$ the power of the motor increases to approximately 112 HP!

## 3. Conclusion

We have studied the possibility to control the gravity on luminescent materials and have concluded that *eletroluminescent materials with high refractive indices* are a new and efficient solution for the gravity control technology. Particularly in the case of the gravitational motors.

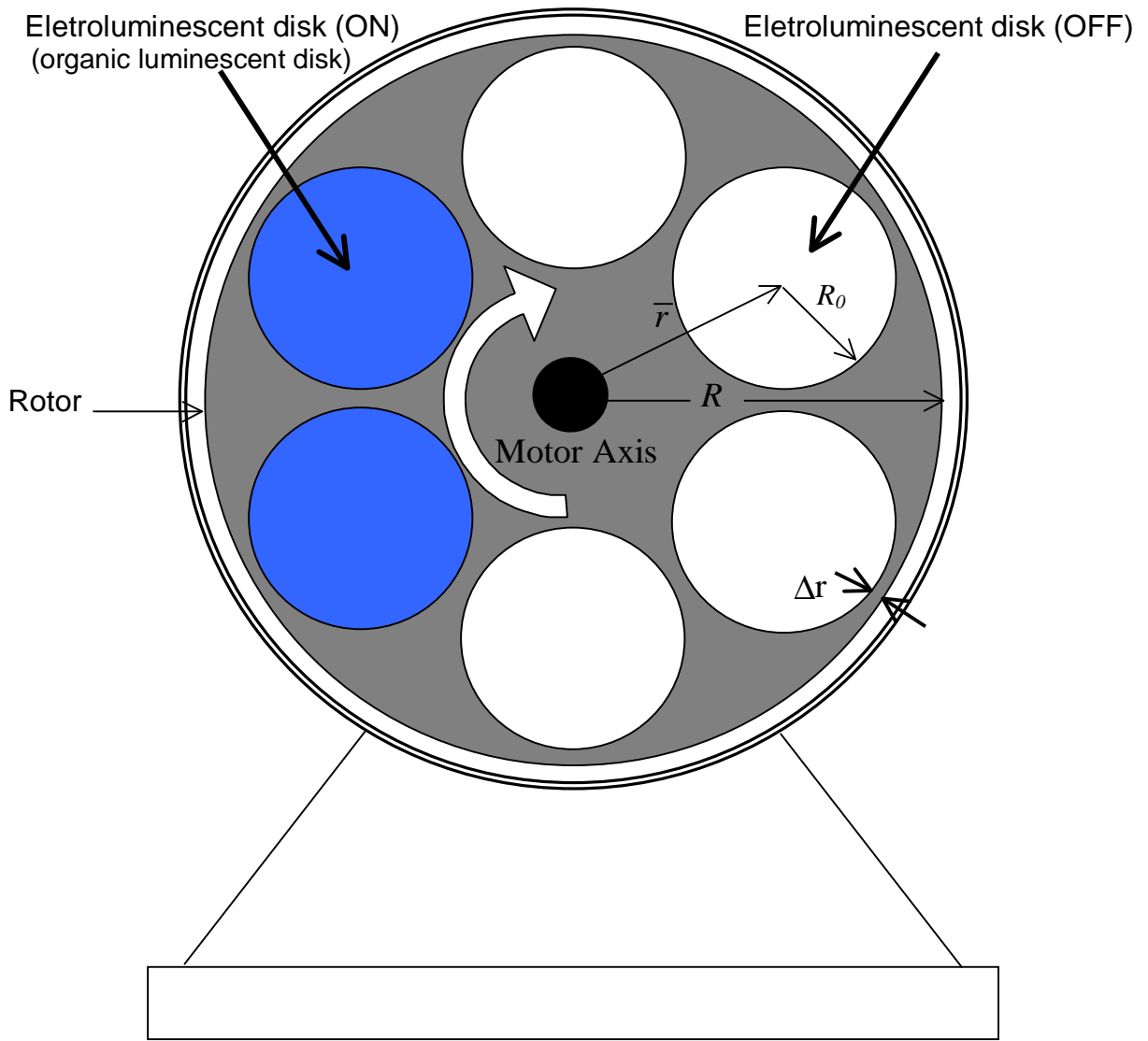

(a) Cross-section of the Motor

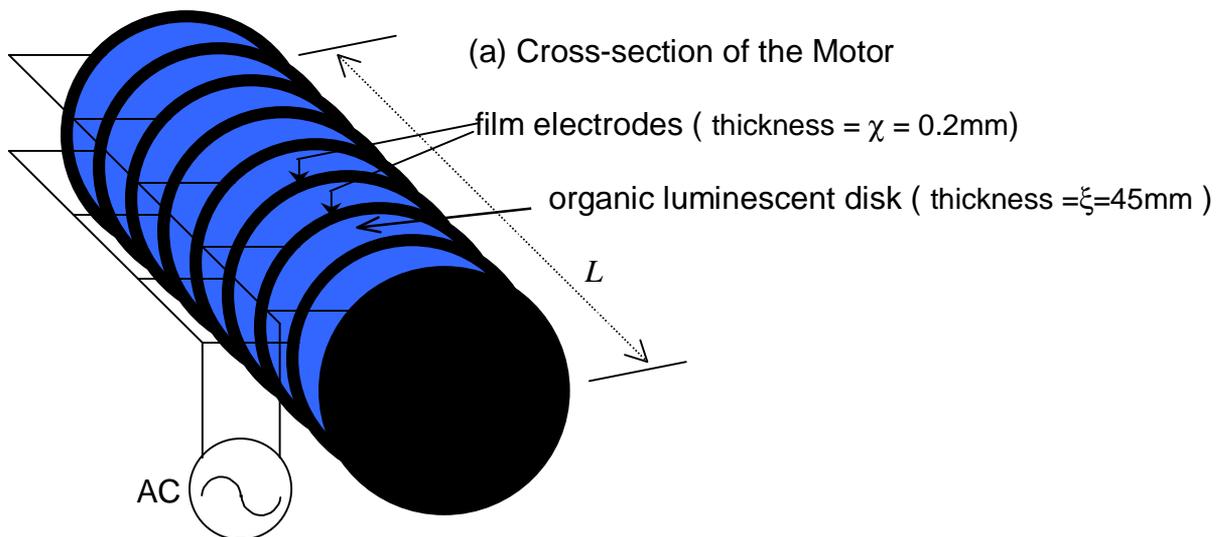

(b) Schematic diagram of the battery of (EL)cells

Fig.1 – The Gravitational Motor

# APPENDIX A

It is important to note that the *momentum* $q$ in Eq.(1) can be also produced by an Electric and/or Magnetic field if the particle has an electric charge $Q$.

In that case, combination of *Lorentz's Equation* $\vec{F} = Q\vec{E}_0 + Q\vec{V} \times \vec{B}$ and $\vec{F} = m_g \vec{a}$ (see reference 1, p.78-Eq.(2.05)) gives

$$q = m_g V = m_g \frac{Q(\vec{E}_0 + \vec{V} \times \vec{B})}{m_g} \Delta t \qquad (A1)$$

In the particular case of an oscillating particle ( frequency $f, \Delta t = 1/f$ ) we have

$$q = \frac{Q(\vec{E}_0 + \vec{V} \times \vec{B})}{f} \qquad (A2)$$

Let us consider a *parallel-plate capacitor* where $d$ is the distance between the plates; $\Delta V$ is the applied voltage; $E_0 = \Delta V / d$ is the external electric field. Inside the dielectric the electric field is $E = \sigma/\varepsilon = E_0/\varepsilon_r$ where $\sigma$ (in C/m²) is the density of electric charge and $\varepsilon = \varepsilon_r \varepsilon_0$.

Thus the charge $Q$ on each surface of the dielectric is given by $Q = \sigma S$ ( $S$ is the area of the surface). Then we have

$$Q = \sigma S = (E\varepsilon)S = (E\varepsilon_r \varepsilon_0)S = E_0 \varepsilon_0 S \qquad (A3)$$

Within the field $E_0$, the charge $Q$ (or "charge layer") acquire a *momentum* $q$, according to Eq.(A2), given by

$$q = \frac{QE_0}{f} = \frac{E_0^2 \varepsilon_0 S}{f} = \frac{(\Delta V / d)^2 \varepsilon_0 S}{f} \qquad (A4)$$

Assuming that in the dielectric of the capacitor there is $N^*$ *layers of dipoles* with thickness $\xi$ approximately equal to the diameter of the atoms ,i.e., $N^* = d/\xi \cong 10^{10} d$ then, according to Eq.(1), for $q \gg m_i c$, the gravitational mass $m_g^*$ of each *dipole layer* is

$$m_g^* \cong -2\left(\frac{q}{m_i c}\right) m_i \cong -\frac{2q}{c} \cong$$

$$\cong -2\left(\frac{\Delta V}{d}\right)^2 \frac{\varepsilon_0 S}{fc} \qquad (A5)$$

Thus, the total gravitational mass $m_g$ of the dielectric may be written in the form

$$m_g = N^* m_g^* \cong -2 \times 10^{10} \left(\frac{\varepsilon_0 S}{fcd}\right) \Delta V^2 \qquad (A6)$$

For example, if we have $\Delta V = 50KV; S = 0.01 m^2; f \cong 10^2 Hz; d = 1mm$ Eq.(A6) gives

$$m_g \cong -0.15 kg$$

Possibly this is the explanation for the *Biefeld-Brown Effect*.